# РАЗВИТИЕ ФОРМАЛЬНЫХ МОДЕЛЕЙ, АЛГОРИТМОВ, ПРОЦЕДУР, РАЗРАБОТКИ И ФУНКЦИОНИРОВАНИЯ ПРОГРАММНОЙ СИСТЕМЫ "ИНСТРУМЕНТАЛЬНЫЙ КОМПЛЕКС ОНТОЛОГИЧЕСКОГО НАЗНАЧЕНИЯ"

*А.В. Палагин, Н.Г. Петренко, В.Ю. Величко, К.С. Малахов*

Институт кибернетики имени В.М. Глушкова НАН Украины,
Киев-187 МСП, 03680, проспект Академика Глушкова, 40,
email: palagin_a@ukr.net, факс: +38044 5263348

Рассмотрено обобщенное представление математической модели программной системы ИКОН, разработаны формальные модели ПС ИКОН, представленные в аналитическом виде UML-диаграмм. Описана трехуровневая архитектура ПС ИКОН в среде клиент-сервер, а также процесс разработки сложных программных систем.

The given paper describes a generalized representation of the mathematical model of the software system Instrumental Complex for Ontological Engineering Purpose (ICOEP). ICOEP is a system that implements one of the areas of complex technologies Data & Text Mining, namely, analysis and processing of linguistic corpus in Ukrainian and/or Russian, extraction of subject knowledge from them and representation of knowledge in ontology of the subject domain. To solve the problems of scalability, performance and security, a concept for the three-tier architecture of the ICOEP in the client-server environment was developed. Multitier architecture of the ICOEP consists of a presentation tier, a logic tier and a data tier. Functional-component model and UML diagrams (use case diagram, activity diagram, software entities diagram) of the ICOEP are presented. The process of developing complex software systems is described.

## Введение

На протяжении нескольких десятилетий задача поиска эффективного повторяемого и предсказуемого процесса или методологии, позволяющей улучшить производительность, качество и надежность разработки программного обеспечения (ПО) и программных систем (ПС) является актуальной. Известны работы, в которых предлагается (на базе методов структурного программирования) технология надежной разработки ПО и ПС, используя при этом тестирование и верификацию программ на основе методов доказательного проектирования [1]. Разработка программных систем [2, 3] (Software development process, Software development life-cycle (SDLC)) – это процесс, направленный на создание и поддержку работоспособности, качества и надежности ПС, используя технологии и методологии из информатики, математики и других инженерно-прикладных дисциплин. Полный процесс разработки ПС [2, 4] выполняется в несколько этапов.

1. Определение требований к ПС (Requirements, Specification): извлечение, анализ, спецификация и утверждение требований для ПС (или составление технического задания на проектирование ПС).

2. Проектирование ПС (Architecture, Design): построение формальной модели ПС, проектирование ПС средствами Computer-Aided Software Engineering (описание моделей ПС на унифицированном языке моделирования UML).

3. Инженерия ПС (Software engineering): реализация ПС с помощью выбранного языка программирования (написание исходного кода).

4. Тестирование и отладка (Testing, Debugging): поиск и исправление ошибок в программных модулях ПС.

5. Подготовка документации на ПС (Software documentation): множество необходимых инструкций пользователя ПС.

6. Внедрение программной системы (Deployment): процесс настройки ПС под определенные условия эксплуатации, а также обучение пользователей работе с программным продуктом.

7. Сопровождение программной системы (Maintenance): процесс оптимизации и устранения обнаруженных в процессе опытной эксплуатации дефектов функционирования ПС, передача ПС в эксплуатацию.

За рамками приведенного выше списка этапов остались такие пункты как: выбор методологии процесса разработки ПО и ПС [5–7] (основными парадигмами считаются Agile (Scrum), Waterfall, Rapid application development (RAD), Test driven development (TDD)), локализация ПС. Эти пункты требуют более детального рассмотрения и описания, что выходит за рамки данной работы.

## Постановка задачи

Разработка формальной модели ПС – это важный и основополагающий этап при проектировании сложной ПС, требующий системного анализа выбора математических структур, наиболее подходящих для моделирования заданного набора задач. Разработанная формальная модель должна позволять выполнять:

– многокритериальные качественные и количественные оценки работы системы;

– оптимизацию алгоритмов функционирования моделей на протяжении всего жизненного цикла системы, в соответствии с критериями сложности реализации ПС и соответствующих алгоритмов, производительности, ограничения реального времени получения результата и другие.

С учетом изложенной выше концепции создания программных систем, необходимо разработать формальные модели инструментального средства автоматизированного построения онтологий предметных областей, названного "Инструментальным комплексом онтологического назначения" (ИКОН).

## Программная система "Инструментальный комплекс онтологического назначения"

ИКОН является системой, реализующей одно из направлений комплексных технологий Data & Text Mining, а именно – анализ и обработку больших объемов неструктурированных данных, в частности лингвистических корпусов текстов на украинском и/или русском языке, извлечение из них предметных знаний с последующим их представлением в виде системно-онтологической структуры или онтологии предметной области (ПдО). ИКОН предназначен для реализации множества компонентов интегрированной информационной технологии [8].

Современный этап развития и построения программных систем характеризуется существенным усложнением процесса их разработки. Под программной системой понимается некоторая конечная совокупность программ, предназначенных для достижения поставленной цели (целей). Функционирование системы заключается в использовании входящих в нее программ (программных модулей), и может быть выполнено двумя способами [9]:

- однопотоковое выполнение процессов (в каждый момент времени выполняется процедура (оператор) только в одной из программ, либо выполняется только одна программа (модуль) и все остальные программы (модули) в этот момент времени приостанавливают свою работу);
- многопотоковое (мультипрограммное) выполнение процессов (в каждый момент времени выполняется процедура (оператор) в каждой программе, либо параллельно выполняются несколько или более программ).

При построении модели ПС следует руководствоваться следующими принципами [10].

1. Модель системы не должна быть чрезмерно детальной (излишняя сложность модели может вызвать существенные вычислительные проблемы при ее формальном анализе).
2. Модель системы не должна быть чрезмерно упрощенной, она должна отражать те аспекты системы, которые имеют отношение к проверяемым свойствам, и сохранять все свойства моделируемой системы, представляющие интерес для анализа.

С точки зрения программной инженерии программная система рассматривается в виде набора описаний, представленных в виде математических моделей, формализмов и техник моделирования [11, 12].

Структура математических моделей ПС такого рода включает в себя следующие модели [11, 12]:

1) информационная модель;
2) функционально-компонентная модель.

Опишем каждую из перечисленных моделей.

## Информационная модель ПС ИКОН

Информационные модели используются для представления и описания потоков информации, структур данных, а также программ (модулей) в программной системе.

Обобщенная информационная модель ПС ИКОН представляется некоторой конечной совокупностью программ (программных модулей) [10–12]:

$$ИКОН = \sum_{i=1}^{n} \Pi_{ИКОН_i}, \qquad (1)$$

где $i = \overline{1, n}$,

$n$ – количество программных модулей входящих в ПС ИКОН,

$\Pi_{ИКОН_i}$ – некоторая программа (модуль) ПС ИКОН.

При этом реализуется отображение $G_{\Pi_{ИКОН}}$ интеграции функций множества программных модулей ИКОН в обобщенную (целевую) функцию ИКОН:

$$G_{\Pi_{ИКОН}} : S_{ИКОН} \to F_{ИКОН}, \qquad (2)$$

где $S_{ИКОН}$ – множество функций (набор функций) определенного программного модуля ИКОН,

$F_{ИКОН}$ – обобщённая функция ИКОН.

На данном этапе жизненного цикла ПО ПС ИКОН можно представить следующей совокупностью программных модулей:

$$\Sigma_{ИКОН} = \{\Pi_{МУГО}, \Pi_{МЛАТД}, \Pi_{МВПОС}, \Pi_{МПТД}, \Pi_{МИПТИ}, \Pi_{МУБДЗ}, \Pi_{МБД}, \Pi_{МПО}\}. \quad (3)$$

Рассмотрим более детально функциональные характеристики программных модулей ИКОН.

1. $\Pi_{МУГО}$ – набор функций программного модуля "Управляющая графическая оболочка",

$$S_{ИКОН}(\Pi_{МУГО}) = < S_1, S_2, S_3, S_4, S_5, S_6 >, \quad (4)$$

где $S_1$ – осуществляет предварительное наполнение среды внешними электронными коллекциями энциклопедических, толковых словарей и тезаурусов, описывающих домен предметных знаний;

$S_2$ – обеспечивает запуск и последовательность исполнения прикладных программ, реализующих составные информационные технологии проектирования онтологии ПдО и системной интеграции междисциплинарных знаний;

$S_3$ – отображает ход процесса проектирования онтологии ПдО;

$S_4$ – содержит позиции меню для запуска как последовательностей, так и отдельных прикладных программ, используемых в процессе проектирования онтологии ПдО;

$S_5$ – обеспечивает интерфейс с естественным интеллектом;

$S_6$ – обеспечивает обмен информацией между прикладными программами и базами данных.

2. $\Pi_{МПТД}$ – набор функций программного модуля поиска текстовых документов,

$$S_{ИКОН}(\Pi_{МПТД}) = < S_7, S_8 >, \quad (5)$$

где $S_7$ – поиск текстовой информации во внешних источниках;

$S_8$ – формирование лингвистического корпуса текстов в библиотеке (базе данных) текстовых документов.

3. $\Pi_{МИПТИ}$ – набор функций программного модуля индексации и поиска текстовой информации,

$$S_{ИКОН}(\Pi_{МИПТИ}) = < S_9 >, \quad (6)$$

где $S_9$ – создание и манипулирование индексами, по которым возможно найти информацию в соответствующих библиотеках ТД, тезаурусах, онтологиях.

4. $\Pi_{МЛАТД}$ – набор функций программного модуля лингвистического анализа текстовых документов,

$$S_{ИКОН}(\Pi_{МЛАТД}) = < S_{10}, S_{11}, S_{12} >, \quad (7)$$

где $S_{10}$ – формирование множества терминов;

$S_{11}$ – формирование множества понятий;

$S_{12}$ – формирование множества отношений между понятиями;

5. $\Pi_{МПО}$ – набор функций программного модуля построения онтологии ПдО,

$$S_{ИКОН}(\Pi_{МПО}) = < S_{13}, S_{14}, S_{15}, S_{16}, S_{17}, S_{18}, S_{19} >, \quad (8)$$

где $S_{13}$ – считывает из модуля визуального проектирования онтологических структур начальную онтологию ПдО. Из библиотеки ЭК считывает определения понятий, входящих в начальную онтологию, и формирует множества функций интерпретации;

$S_{14}$ – считывает из соответствующей библиотеки онтографы и формализованные описания онтологий ТД, проверяет их на непротиворечивость;

$S_{15}$ – анализирует на полноту множества функций интерпретации. Просматривает электронные коллекции энциклопедических, толковых словарей и тезаурусов и пополняет множества последних для соответствующих понятий-объектов и понятий-процессов;

$S_{16}$ – в процессе построения онтологии, по необходимости, обращается к любому информационному хранилищу;

$S_{17}$ – выполняет поочередную привязку онтографов и функций интерпретации онтологий ТД;

$S_{18}$ – выполняет привязку общего онтографа ТД и множеств функций интерпретации с онтографом и множествами функций интерпретации начальной онтологии соответственно;

$S_{19}$ – передает полученный результат в модуль визуального проектирования на верификацию.

6. $\Pi_{МВПОС}$ – набор функций программного модуля визуального проектирования онтологических структур,

$$S_{ИКОН}(\Pi_{МВПОС}) =< S_{20}, S_{21}, S_{22} >, \quad (9)$$

где $S_{20}$ – разработка начальной онтологии ПдО;

$S_{21}$ – ручное неавтоматизированное проектирование онтологии ПдО;

$S_{22}$ – верификация онтографа.

7. $\Pi_{МУБДЗ}$ – набор функций программного модуля управления библиотеками данных и знаний,

$$S_{ИКОН}(\Pi_{МУБДЗ}) =< S_{23}, S_{24} >, \quad (10)$$

где $S_{23}$ – осуществляет запись, накопление и хранение информации в соответствующих библиотеках;

$S_{24}$ – по запросу менеджера проектов осуществляет чтение из библиотек запрашиваемой информации.

8. $\Pi_{МБД}$ – набор функций программного модуля базы данных,

$$S_{ИКОН}(\Pi_{МБД}) =< S_{25}, S_{26}, S_{27}, S_{28}, S_{29}, S_{30}, S_{31} >, \quad (11)$$

где $S_{25}$ – накопление и хранение онтологий ПдО;

$S_{26}$ – накопление и хранение ТД;

$S_{27}$ – накопление и хранение ЛКТ;

$S_{28}$ – накопление и хранение списков множеств терминов, понятий и отношений;

$S_{29}$ – накопление и хранение проектов ИКОН;

$S_{30}$ – поиск по БД;

$S_{31}$ – работа в режиме клиент-сервер.

## Функционально-компонентная модель ПС ИКОН

Функционально-компонентная модель используется для представления взаимодействий, отношений и зависимостей программных модулей в ПС, а также для подробного описания компонентов системы. Обобщенная функционально-компонентная модель ПС ИКОН может быть представлена в виде:

$$S_{ИКОН} =< M_{Dyn}, M_{Stat}, M_{Phys}, M_{Cmp}, P_0(M_{Dyn}, M_{Stat}) > . \quad (12)$$

Рассмотрим более подробно функционально-компонентную модель ПС ИКОН.

1. $M_{Dyn}$ – модель, определяющая поведение системы. Рассматривается как UML dynamic model [13–15]: диаграммы вариантов использования (use case diagram); диаграммы активности (activity diagram), диаграммы взаимодействий (sequence diagram); *Динамическая модель (Dynamic model)* описывает поведение системы, например, изменение программных сущностей (software entities) во время выполнения приложения.

$$M_{Dyn} =< d_{use}, d_{act}, d_{seq} >, \quad (13)$$

где $d_{use}$ – множество UML-диаграмм вариантов использования ПС ИКОН,

$d_{act}$ – множество UML-диаграмм активности ПС ИКОН,

$d_{seq}$ – множество UML-диаграмм взаимодействий ПС ИКОН.

2. $M_{Stat}$ – модель, определяющая структуру системы. Рассматривается как UML static model [13–15]: диаграммы классов (class diagram) и описание основных модулей системы в виде статических блок-схем. *Статическая модель (Static model)* описывает элементы системы и их взаимоотношения (классы, атрибуты, операторы).

$$M_{Stat} =< d_{class}, b_{Stat}, req >, \quad (14)$$

где $d_{class}$ – множество UML-диаграмм классов ПС ИКОН,

$b_{Stat}$ – множество статических блок-схем ПС ИКОН,

$req$ – техническое задание на проектирование ПС ИКОН.

3. $M_{Phys}$ – модель, определяющая структуру программных сущностей. Рассматривается как файлы исходного кода, библиотеки, исполняемые файлы, и отношения между ними. *Физическая модель (Physical model)* [13] отображает неизменяемую структуру программных сущностей, в частности файлов исходного кода, библиотек, исполняемых файлов, и отношений между ними.

$$M_{Phys} = < src, lib, exef >, \qquad (15)$$

где $src$ – файлы исходного кода,

$lib$ – программные библиотеки,

$exef$ – исполняемые файлы.

4. $M_{Cmp}$ – модель (схема) компонентов ПС, используется для визуализации высокоуровневой структуры системы и поведения служб (сервисов), предоставляемых и потребляемых этими элементами (компонентами) через интерфейсы. Компонент – это модульная единица, заменяемая в пределах среды. Его внутренние составляющие скрыты, но доступ к функциям компонента можно получить с помощью одного или нескольких четко определенных предоставленных интерфейсов (*provided interfaces*). Компонент также может иметь требуемые интерфейсы (*required interfaces*). Требуемый интерфейс определяет, какие функции и службы (сервисы) он требует от других компонентов. Объединив предоставленные и требуемые интерфейсы нескольких компонентов, можно создать более крупный компонент. Можно сказать, что вся ПС, по сути, представляет собой компонент [16].

Создание схем компонентов имеет несколько преимуществ.

- Анализ основных элементов ПС позволяет команде разработчиков понять принципы существующей системы и создать новую.
- Представление ПС в качестве коллекции компонентов с четко определенными предоставленными и требуемыми интерфейсами позволяет более эффективно разделить компоненты. В свою очередь, это облегчает понимание конструкции ПС и ее корректирование при изменении требований.
- Использование диаграммы (схемы) компонентов для представления конструкции ПС независимо от того, какой язык программирования или программная платформа используется сейчас или будет использоваться в будущем.

5. $P_0(M_{Dyn}, M_{Stat})$ – предикат целостности системы, определяющий назначение системы, семантику (смысл) моделей $M_{Dyn}$ и $M_{Stat}$ [17];

UML-диаграмма вариантов использования высокого уровня абстракции ПС ИКОН показана на рис. 1.

Суть данной диаграммы состоит в следующем: проектируемая система представляется в виде множества сущностей или акторов (actor), взаимодействующих с системой с помощью так называемых вариантов использования. При этом актором или действующим лицом называется любая сущность, взаимодействующая с системой извне. Это может быть человек, техническое устройство, программа или любая другая система, которая может служить источником воздействия на моделируемую систему. В свою очередь, вариант использования служит для описания сервисов, которые система предоставляет актору.

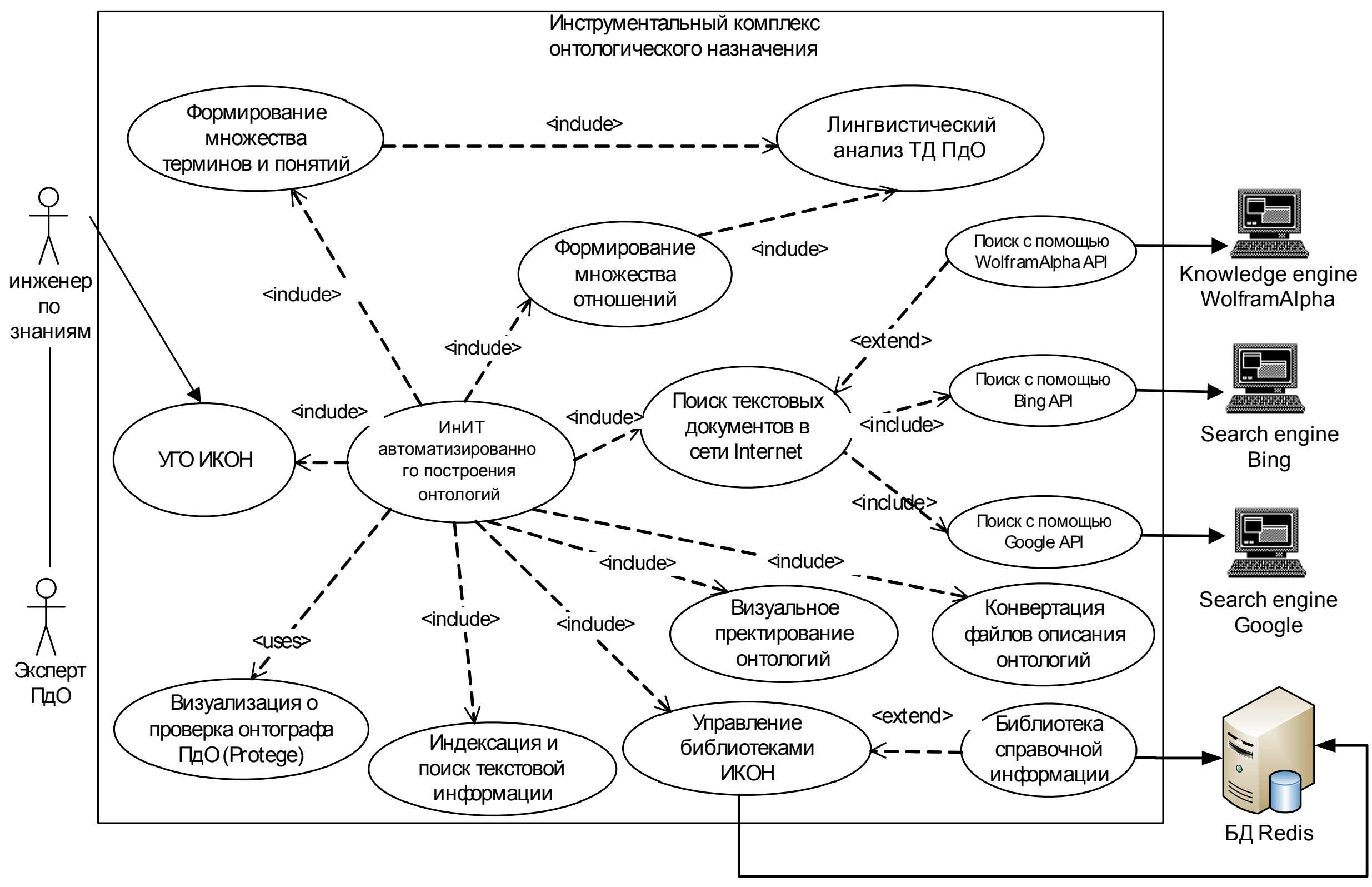

Рис. 1. UML-диаграмма вариантов использования ПС ИКОН.

UML-диаграмма активности высокого уровня абстракции ПС ИКОН показана на рис. 2.

С помощью *диаграммы активности* (рис. 2) можно изучать поведение системы с использованием моделей потока данных и потока управления. Диаграмма активности отображает некоторый алгоритм, описывающий жизненный цикл объекта, состояния которого могут меняться.

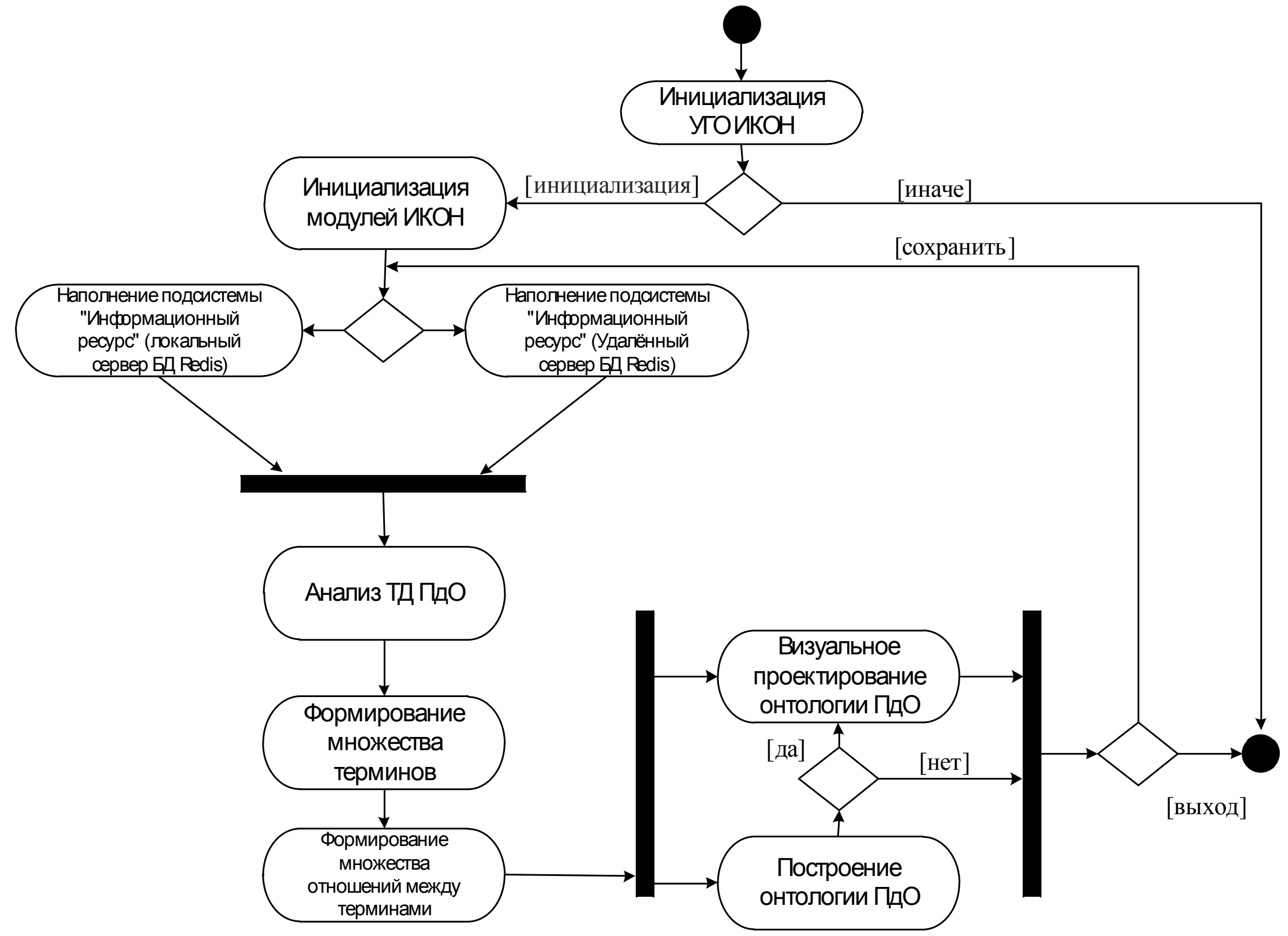

Рис. 2. UML-диаграмма активности ПС ИКОН

Физическая модель программных сущностей ПС ИКОН показана на рис. 3.

Представленная модель отображает неизменяемую структуру программных сущностей, в частности файлов исходного кода, библиотек, исполняемых файлов, и отношений между ними.

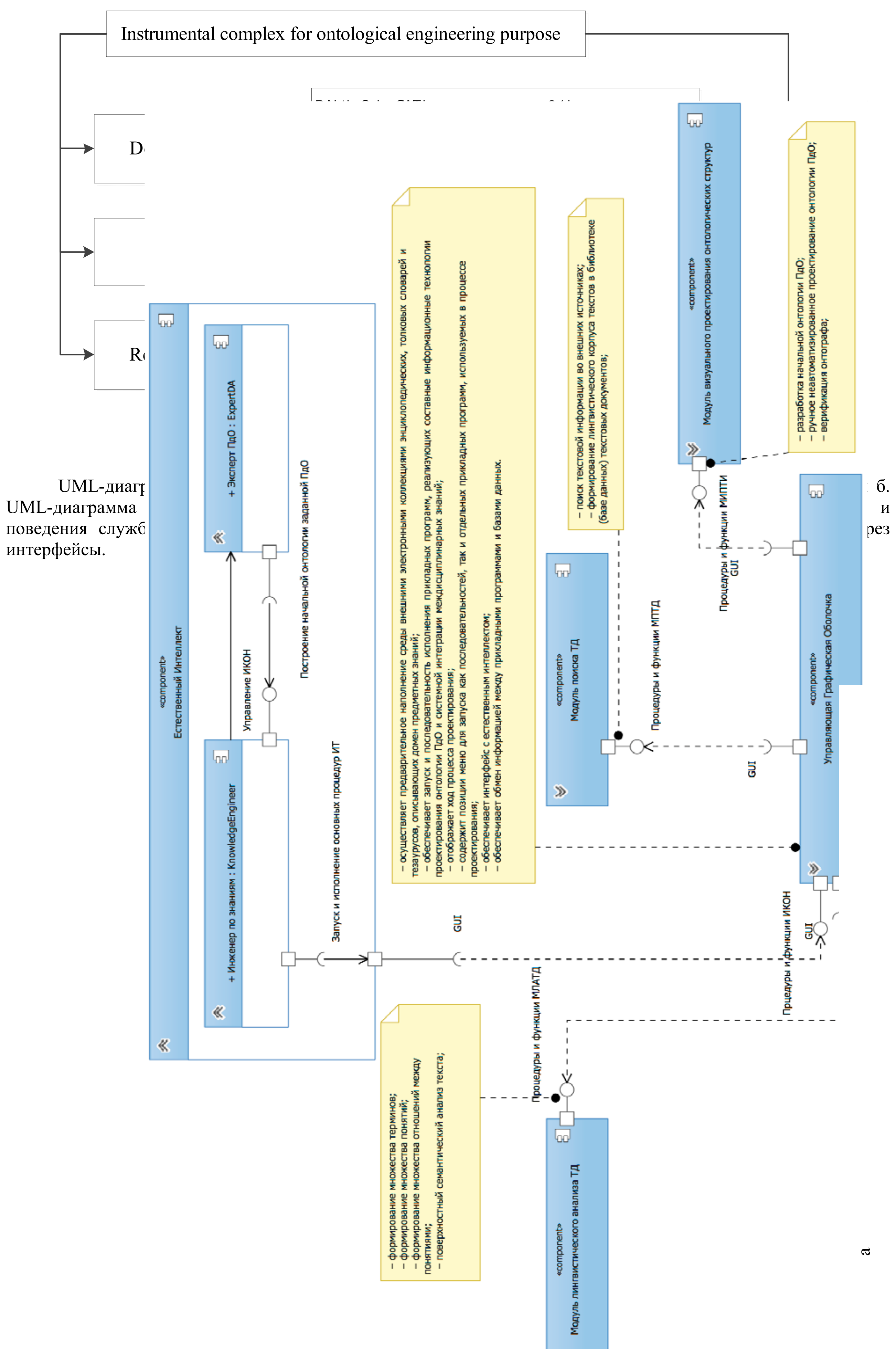

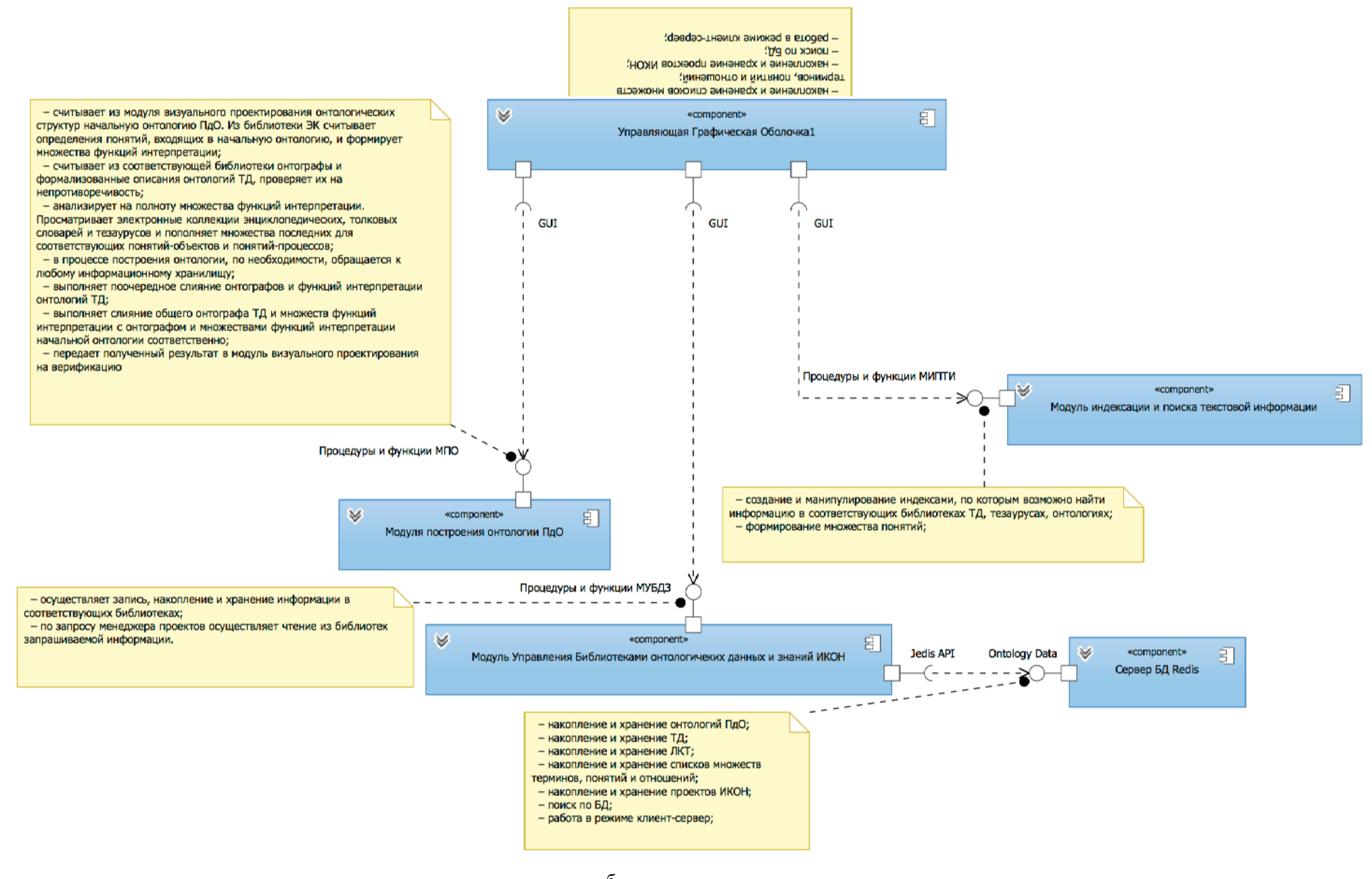

б

Рис. 4. UML-диаграмма компонентов высокого уровня абстракции ПС ИКОН

## Трехуровневая архитектура ПС ИКОН

В процессе разработки ПС ИКОН возник ряд проблем, в частности, проблемы масштабируемости, производительности, безопасности. Для решения вышеуказанных проблем была разработана концепция трёхуровневой архитектуры ПС ИКОН.

В терминах программной инженерии трехуровневая архитектура (Multitier architecture) состоит из уровня представления (Presentation tier), уровня логики (Logic tier), уровня данных (Data tier) и предполагает наличие следующих компонентов приложения: клиентское приложение ("тонкий клиент" или терминал), подключенное к серверу приложений, который в свою очередь подключен к серверу базы данных (БД) [18–20].

Клиент – это интерфейсный (обычно графический) компонент, который реализует уровень представления, собственно приложение для конечного пользователя. Первый уровень не должен иметь прямых связей с базой данных (по требованиям безопасности), быть нагруженным основной бизнес-логикой [21] (по требованиям масштабируемости) и хранить состояние приложения (по требованиям надежности). На первый уровень может быть вынесена и обычно выносится простейшая бизнес-логика: интерфейс авторизации, алгоритмы шифрования, проверка вводимых значений на допустимость и соответствие формату, несложные операции (сортировка, группировка, подсчет значений) с данными, уже загруженными на терминал.

Сервер приложений реализует уровень логики ПС ИКОН, на котором сосредоточена бо́льшая часть бизнес-логики (описание функциональных алгоритмов, обрабатывающих обмен информацией между сервером БД и клиентом). Вне его остаются фрагменты, экспортируемые на терминалы, а также погруженные в третий уровень хранимые процедуры и функции.

Сервер БД обеспечивает хранение данных и реализует третий уровень – уровень данных. Обычно это стандартная реляционная или объектно-ориентированная система управления базами данных (СУБД). Если третий уровень представляет собой БД вместе с хранимыми процедурами и схемой, описывающей приложение в терминах реляционной модели, то второй уровень строится как программный интерфейс, связывающий клиентские компоненты с прикладной логикой (программным модулем) БД.

В простой конфигурации физически сервер приложений может быть совмещен с сервером БД на одном компьютере, к которому по сети подключается один или несколько терминалов. В "правильной" (с точки зрения безопасности, надежности, масштабируемости) конфигурации сервер БД находится на выделенном компьютере (или кластере), к которому по сети подключены один или несколько серверов приложений, к которым, в свою очередь, по сети подключаются терминалы (рис. 5).

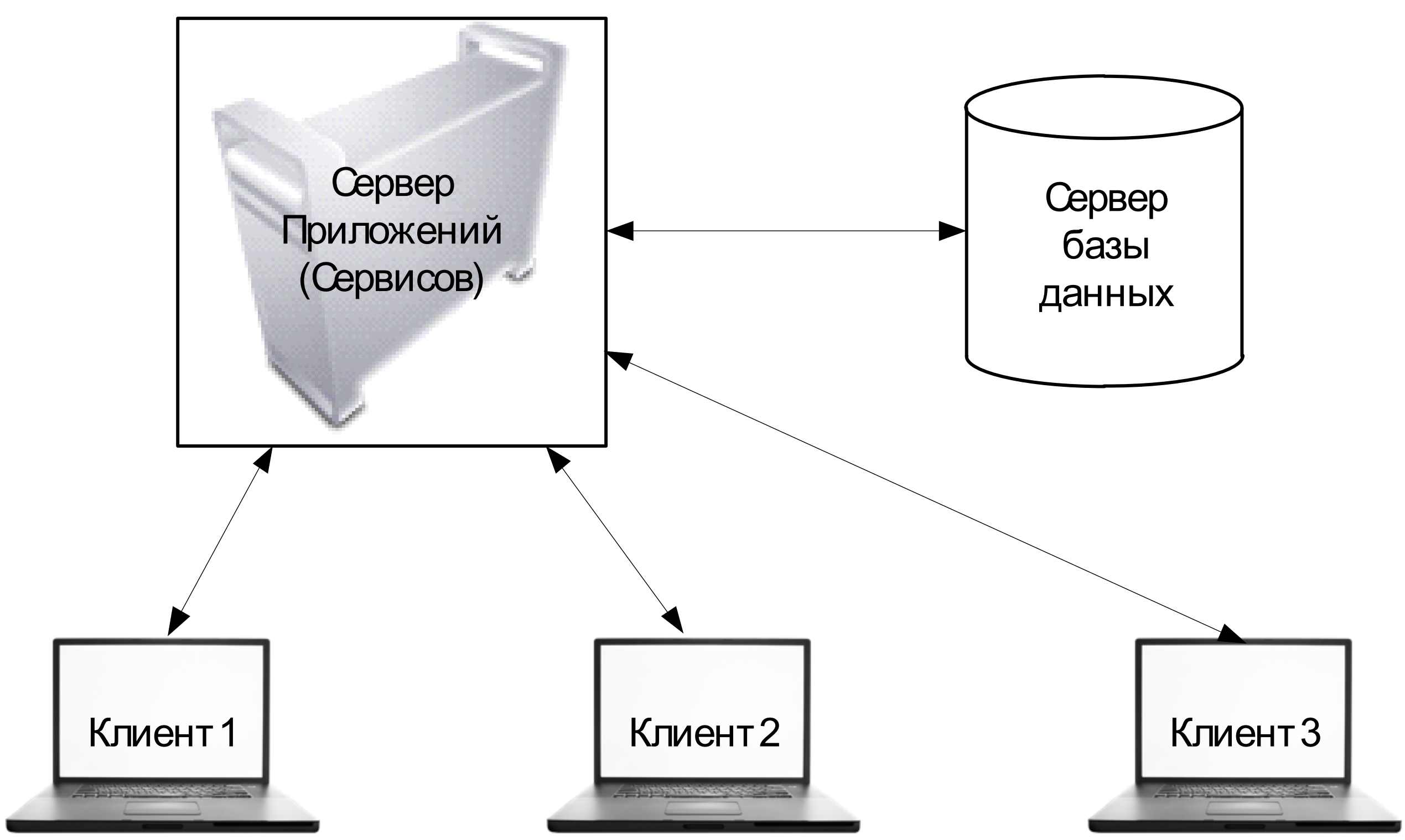

Рис. 5. Пример трехуровневой архитектуры

Основные достоинства трехуровневой архитектуры ПС:

– масштабируемость – способность системы справляться с увеличением рабочей нагрузки (увеличивать свою производительность) при добавлении аппаратных ресурсов;

– конфигурируемость – изолированность уровней друг от друга позволяет (при правильном развертывании архитектуры) быстро и простыми средствами переконфигурировать систему при возникновении сбоев или при плановом обслуживании на одном из уровней;

– высокие безопасность и надежность;

– низкие требования к скорости передачи информации канала (сети) между терминалами и сервером приложений;

– низкие требования к производительности и техническим характеристикам терминалов, как следствие снижение их стоимости. Терминалом может выступать не только компьютер, но и, например, мобильный телефон.

К недостаткам трехуровневой архитектуры ПС в сравнении с файл-серверной архитектурой следует отнести:

– более высокая сложность создания приложений;

– сложнее в разворачивании и администрировании;

– высокие требования к производительности серверов приложений и сервера базы данных, а, значит, и высокая стоимость серверного оборудования;

– высокие требования к скорости канала (сети) между сервером базы данных и серверами приложений.

Рассмотрим реализацию трехуровневой архитектуры ИКОН. На рис. 6 представлена обобщенная структурная схема трехуровневой архитектуры ИКОН.

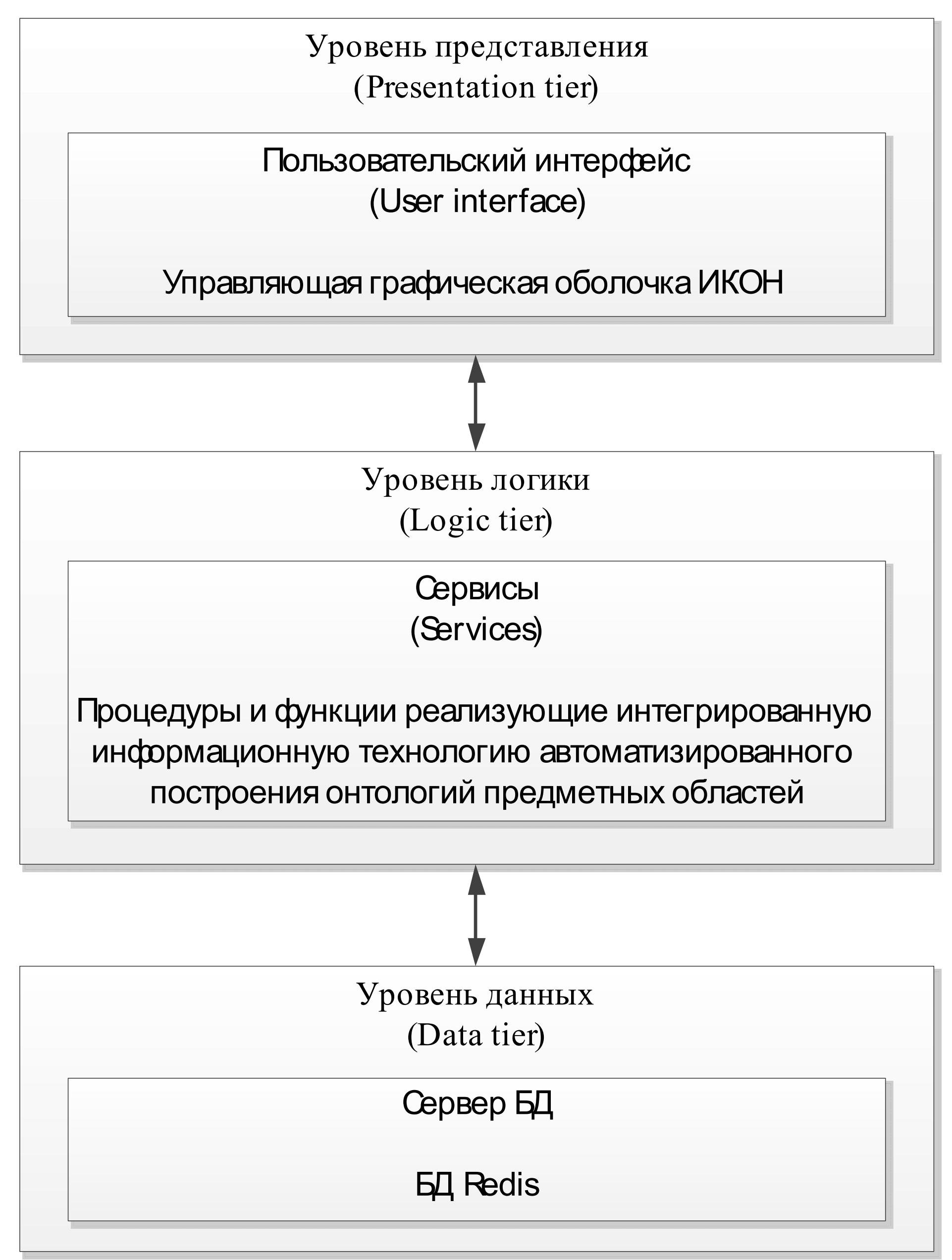

Рис. 6. Обобщенная структурная схема трехуровневой архитектуры ИКОН

На уровне представления реализовано клиентское приложение "Управляющая графическая оболочка". На уровне логики реализованы процедуры и функции, обеспечивающие функционирование программных модулей ПС ИКОН и предоставляемых ими сервисов (модуль управления библиотеками; модуль индексации и поиска текстовой информации; модуль построения онтологии ПдО; модуля поиска текстовых документов; модуля лингвистического анализа текстовых документов; модуля визуального проектирования онтологических структур; модуль построения отношений между понятиями). На уровне данных реализован сервер БД Redis, обеспечивающий хранение данных ИКОН (лингвистического корпуса текстов, цифровой библиотеки справочной информации, онтологических структур, терминов, понятий, словарей, тезаурусов, текстовых документов). Уровни архитектуры ПС ИКОН находятся в двусторонней зависимости, которая указывает на то, что один уровень может использовать функции другого уровня и наоборот.

Блок-схема алгоритма работы ПС ИКОН в трехуровневой архитектуре показана на рис. 7.

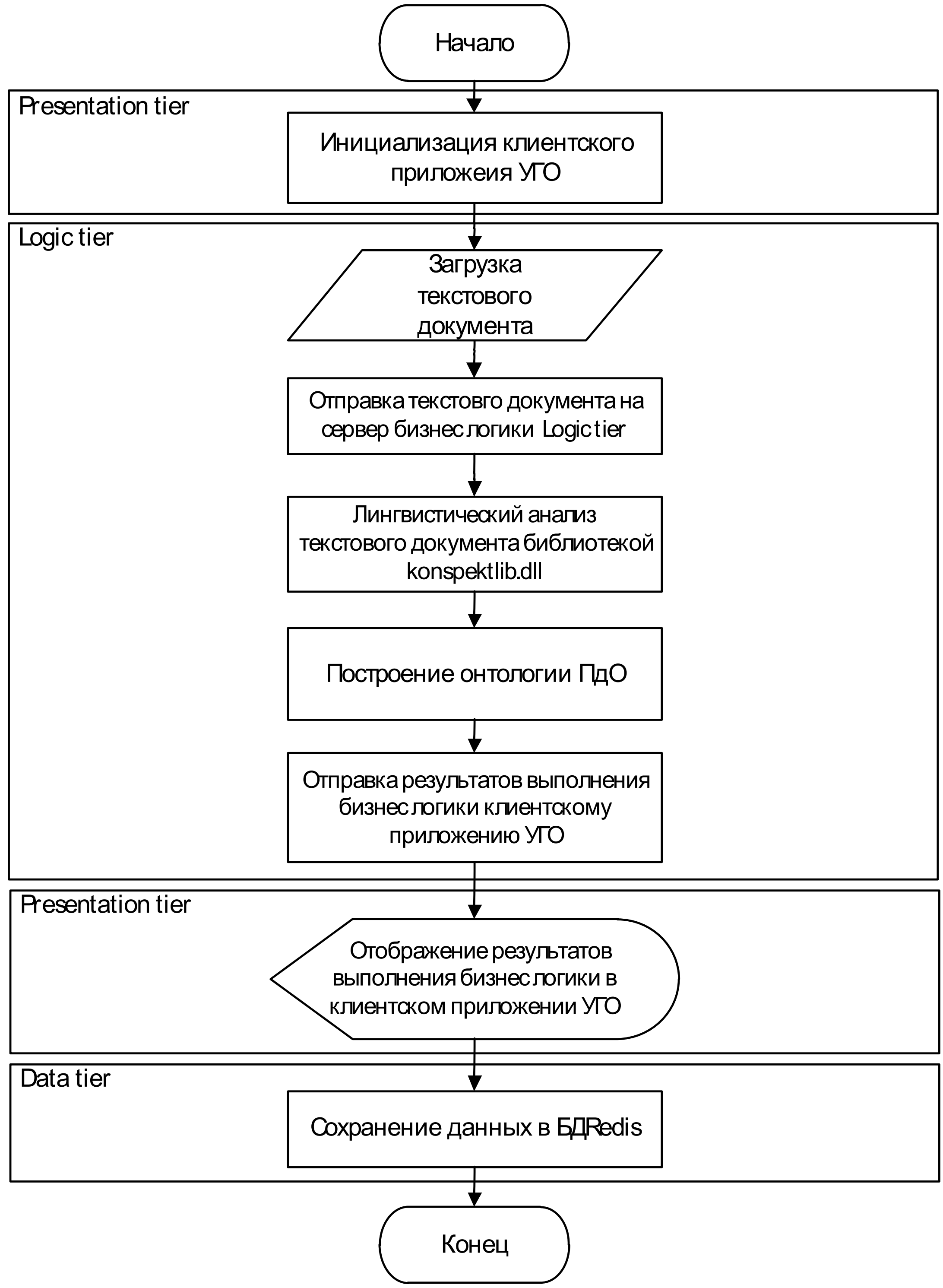

Рис. 7. Алгоритм работы ИКОН в трехуровневой архитектуре

## Выводы

В работе предложено обобщённое представление формальной модели программной системы "Инструментальный комплекс онтологического назначения" автоматизированного построения онтологий предметных областей. Разработаны и дополнены формальные модели ПС ИКОН, в соответствии с развитием алгоритмов, процедур, разработки и функционирования программной системы в целом. Спроектирована функционально-компонентная модель ПС ИКОН, в частности, UML-диаграмма компонентов ПС ИКОН, описывающая взаимодействие и отношения программных модулей в системе. Рассмотрена концепция трёхуровневой архитектуры ПС ИКОН в среде клиент-сервер. Описан полный процесс разработки сложных программных систем. За рамками данной работы остались особенности реализации концепции трёхуровневой архитектуры ПС ИКОН, а также описание методологии разработки ПО и ПС в целом.